\documentclass[11pt]{article}
\usepackage{geometry}
\usepackage{cite}
\usepackage{amsmath}
\usepackage{amssymb}
\usepackage{amsfonts}
\usepackage{graphicx}
\usepackage{caption}
\usepackage{subcaption}
\usepackage{float}
\usepackage[usenames,dvipsnames]{xcolor}
\usepackage{tikz}
\usepackage{pstricks}
\usepackage{lipsum}
\usepackage{appendix}
\usepackage{authblk}
\usepackage{verbatim}
\usepackage{setspace}
\usepackage{gensymb}
\renewenvironment{thebibliography}[1]{
  \begin{oldthebibliography}{#1}
    \setlength{\itemsep}{0em}
    \setlength{\parskip}{0em}
}
{
  \end{oldthebibliography}
  }
\begin{document}
\date{}
\author{Sandip Chowdhury \thanks{sandipc@iitk.ac.in}}
\author{Kunal Pal \thanks{kunalpal@iitk.ac.in}}
\author{Kuntal Pal \thanks{kuntal@iitk.ac.in}}
\author{Tapobrata Sarkar \thanks{tapo@iitk.ac.in}}
\affil{Department of Physics, \\ Indian Institute of Technology, \\ Kanpur 208016, India}
\title{Disformal transformations and the motion of a particle in semi-classical gravity}
\maketitle

\begin{abstract}
	
The approach to incorporate quantum effects in gravity by replacing free particle geodesics
with Bohmian non-geodesic trajectories
has an equivalent description in terms of a conformally related geometry, where the motion is force 
free, with the quantum effects inside the conformal factor, i.e., in the geometry itself. 
For more general disformal transformations relating gravitational and physical geometries,
we show how to establish this equivalence by taking the quantum effects inside the disformal degrees of freedom. 
We also show how one can solve the usual problems associated with the conformal version, namely 
the wrong continuity equation, indefiniteness of the quantum mass, and wrong description of massless particles in the 
singularity resolution argument, by using appropriate disformal transformations.  
\end{abstract}
	
\section{Introduction}
More than twenty five years ago, Bekenstein \cite{JBe} showed that in theories of gravity where
two distinct geometries are present, they are, in general, related by a disformal transformation, which is
a generalization of the conformal transformation. In such situations, the gravitational dynamics is
controlled by the metric and is called gravitational geometry, whereas matter dynamics takes place 
on a geometry that is disformally related to the metric, and called physical geometry. 
This is a notable departure from general relativity (GR), where the dynamics of both gravity and matter 
are determined by the metric, but is common in scalar-tensor theories of gravity, such as 
Brans-Dicke theory, in which the two geometries are related by a conformal transformation. 
The purpose of this paper is to establish the nature of disformal transformations in the context of 
Bohmian mechanics in gravitational backgrounds, and to show that it solves a few important problems that arise in 
an usual treatment popular in the literature, that uses conformal transformations instead. 
	
Bohmian mechanics \cite{BH,PH,DT} is an important tool in a semi-classical understanding of the full theory 
of quantum gravity \cite{SG}. Formulating such a quantum theory of gravity is of course a formidable challenge, 
with the metric playing the role of a quantum operator, and with quantization conditions on space and time. The somewhat
simpler semi-classical approach, in which the metric is classical, has been popular over decades. In Bohmian mechanics,
the particle trajectories are determined by suitable wavefunctions, and the statistical distribution of particle positions 
is given by the modulus squared of this wavefunction. 
In this first quantized approach, one replaces the geodesic motion of freely falling particles in a curved 
space-time by corresponding Bohmian trajectories. In such situations, geodesic equations are 
typically modified by an additional force term coming from a quantum potential \cite{SD,AD,AKh,RG}. This kind of reasoning has been used to deal with the usual singularity problem of classical general relativity \cite{SD}.
	
Importantly, it is known that there is a close relationship between conformal transformations and the motion of a particle 
in Bohmian quantum mechanics. Namely, if we consider the quantum motion of a particle in a flat background,
then this is equivalent to classical motion (one where the effect of quantum potential is absent) on a curved  background which is
related to the previous one by a conformal transformation. The conformal factor is a function of the 
modulus of the wavefunction, and is hence related to the quantum potential (see for e.g. \cite{RG,BK,RC,SS2}  
and references therein).
	
To understand this equivalence beyond the existing literature, in the first part of this work, we consider a 
Klein-Gordon type field, which is governed 
by Bohmian mechanics, and is non-minimally coupled to gravity. After deriving the Raychaudhuri equation, we  
discuss how we can transform to a conformally related frame (with the conformal factor being a variable particle mass) 
where the motion of the particle is force free, but the corresponding field equations have the information of the quantum 
nature of the particle. We argue how to deal with the singularity problem in this frame where the quantum force is 
absent by working out a particular example of such transformation with a suitable choice of the wave function, in 
a Schwarzschild background. 
	
In this context, an important  question to ask is the following. 
If the physical geometry on which the particle moves is different from the gravitational geometry, can we 
perform an analysis similar to the one described above, where the particle, moving in the physical geometry, 
is a quantum mechanical particle obeying Bohmian mechanics ? Is it possible to incorporate quantum 
effects in the disformal factor as in the conformal case? We show it is possible, by  
deriving  the relation between the  acceleration equations and then relating the factors of the transformation 
with quantum potential. 

Next, in a minimally coupled scenario, 
we show that by using the equivalent  description in the disformal frame, we can solve some well known problems 
that are present in the conformal version of theory described before.  For example, in presence of the quantum potential,
the mass of a particle is modified to a variable one (known as the quantum mass of the particle), and this is 
not always positive definite, so that the theory may have tachyonic modes \cite{PH}. If we transform to a 
conformal  frame where the quantum potential is absent, the particle mass becomes constant, but 
the transformation itself is not definite for such imaginary quantum mass. Furthermore the continuity equation 
in the transformed frame does not have the desired form to define a suitable probability density\cite{JCa,TT}. 
One of the main contributions of this paper is to show that instead of transforming 
to a conformal frame, if we use a disformal transformation, all of  the above mention problems can be solved. 
The modified geodesic equation and the continuity equation fix the transformation factors  such that have  positive 
definite values even for a particle with negative quantum mass.
	
In the last part this work we address a related question. If the force on the particle due to it's quantum nature 
is absent in the conformal frame, what happens for a masslss particle moving in null trajectory in that frame. It is 
known that, by the nature of construction of conformal transformation the null geodesics remains invariant, they do 
not feel any force in any of the two conformally related frames. In particular what happens to the argument 
used to avoid the singularity problem. Namely, how can null trajectories not cross if they do not feel any quantum force? 
We shall show how to answer this question  by using the fact that under a general disformal transformation, 
a null trajectory of one frame is not null in another frame \cite{JBe,CCY}. By doing a  disformal transformation 
in a direction different from the wave vector of the photon, we show that photon does not moves in a null trajectory  
in the gravitational geometry, and hence can avoid the singularity. 
	
This paper is organized as follows. In section \ref{sec2}, we briefly review the 
non-geodesic motion of the Bohmian particle in curved spacetime. Then we elaborate how the relation between 
the motion of this particle on a fixed background geometry  can be equivalently described by a force free motion in a 
conformal frame, where the quantum effects are hidden in the energy momentum tensor. In section \ref{sec3}, 
by assuming a minimal coupling between wavefunction and  background geometry we describe how the previous 
set up may be generalized to disformally related spacetimes. Finally, in section \ref{sec4}, we demonstrate how 
one can solve the usual problems that appear in a conformally transformed frame 
by using disformal transformations. The paper ends with conclusions in section \ref{sec5}. 
Throughout this paper, we work in natural units and set $G=c=\hbar=1$.
	
\section{Bohmian motion on a classical background}
\label{sec2}
	
\subsection{The action and the Klein-Gordon equation}\label{kgequation}
	
We consider a quantum particle moving on a timelike path in a fixed classical background. 
The normalized wave function of such a particle $\Psi(x^{\mu})$ can be written in terms of two single valued real 
functions $\mathcal{R}(x^{\mu})$ and $\mathcal{S}(x^{\mu})$, which are the modulus and the phase function respectively,  
as $\Psi(x^{\mu})=\mathcal{R}(x^{\mu})e^{i\mathcal{S}(x^{\mu})}$ \cite{PH,DT}. 
Substituting this form of  wave function in the Schrodinger equation, and separating the real and imaginary parts, one 
can show that, with the four-momentum associated with a particle of mass $m$ guided 
by this wave function  given by $p_{\mu}=\partial_{\mu}\mathcal{S}(x^{\mu})$, 
these two equations can be interpreted as a quantum Hamilton-Jacobi equation, and a continuity equation 
with probability density $\rho=\mathcal{R}^{2}$ \cite{BH,PH}, respectively. But the Schrodinger equation is not  a
manifestly covariant equation.  Thus to write down the relativistic version of the quantum 
Hamilton - Jacobi  equation and the continuity equation, the usual approach is to instead
consider a Klein-Gordon type equation of a particle  (whose dynamics is governed by the 
Bohmian mechanics) described  by the wavefunction  $\Phi(x^{\mu})$ which is assumed to to be 
moving on a fixed classical curved background ($g_{\mu\nu}$)\footnote{ Here   the  field $\Phi(x^{\mu})$ 
(a scalar function) is the first quantized wavefunction of the particle. In this paper no second quantization 
is imposed upon  $\Phi(x^{\mu})$. Nevertheless, we shall use both the words wavefunction and field to 
denote $\Phi(x^{\mu})$.}. 
	
For convenience we start with the following  action, written in standard form, with a non-minimal coupling 
term of $\Phi$ as (see for example \cite{carroll},\cite{RW}):
\begin{equation}\label{EHA}
S_{fR}+S_\Phi+	S_{m}= \int d^{4}x \sqrt{-g}F(\Phi)R +\int  d^{4}x \sqrt{-g} \Big[-\frac{1}{2}
g_{\mu\nu}(\nabla^\mu \Phi) (\nabla^\nu \Phi)  - \frac{1}{2}m^{2}  \Phi^{2} \Big]+\int  d^{4}x \sqrt{-g} 
\mathcal{L}_{m}(g^{\mu\nu}, \lambda_{i})~.
\end{equation}
Here the subscripts in the terms on the left hand side denote the actions due to the non-minimal coupling, the scalar
field and the matter fields (collectively denoted as $\lambda^i$) respectively. 
The coupling between the Ricci scalar $R$, and the scalar field $\Phi$ is assumed  to be  independent of whether it is in the
quantum regime or can be approximately taken to be classical, i.e., independent of the energy scale of the system.

From now on, we shall concentrate on a particular choice of the function $F(\Phi)$, namely
 $F(\Phi)=-\epsilon\Phi^2$. In that case, the gravitational dynamics is governed by the field equation \cite{carroll,IQ}
\begin{equation}\label{EEJF}
G_{\mu\nu}=\frac{1}
{\Phi^{2}}\Big(\nabla_{\mu}\nabla_{\nu}\Phi^{2} - g_{\mu\nu}\Box \Phi^{2}\Big)-\frac{1}{2\epsilon \Phi^{2}}\Big(T^{m}_{\mu\nu}+T^{\Phi}_{\mu\nu}\Big)~,~~\text{with}~~
T^{i}_{\mu\nu}=-\frac{2}{\sqrt{-g}}\frac{\delta S_{i}}{\delta g^{\mu\nu}}~, 
\end{equation}
where $	G_{\mu\nu}$ is the Einstein tensor constructed from the metric $	g_{\mu\nu}$, and  
$T^{m}_{\mu\nu}$ and $T^{\Phi}_{\mu\nu}$ are  the energy momentum tensors associated with matter fields 
and the scalar field respectively. The scalar field dynamics is governed the following Klein-Gordon like equation 
\begin{equation}\label{SEJF}
\big[\Box -m^{2} - \epsilon R \big] \Phi =0~.
\end{equation}
In four spacetime dimensions, this equation is conformally invariant only when the mass term is set to zero 
and  $\epsilon=\frac{1}{6}$ \cite{dab}.  
It is also important to notice that in \cite{SD}, it was assumed that the background metric is non-dynamical in nature 
as a first approximation, i.e., backreaction was neglected. But in the general case with a non zero 
coupling between curvature and the scalar field $\Phi$, this is not possible, unless of course we choose, 
the scalar field to be not dynamical (this fact should  be clear from the field eq.(\ref{EEJF}) above). 
	
Substituting the polar form of the wavefunction $\Phi=\mathcal{R}(x^{\mu})e^{i\mathcal{S}(x^{\mu})}$ in 
the Klein-Gordon equation, and separating real and 
imaginary parts, we get the following  equations
\begin{equation}\label{kgri}
\big(\nabla_{\mu}\mathcal{S}\big)\big(\nabla^{\mu}\mathcal{S}\big)=
-m^{2}-\epsilon R+\frac{\Box\mathcal{R}}{\mathcal{R}}~~~ \text{and} ~~ \nabla_{\mu}\mathcal{J}^{\mu}=0 ~~\text{with}~~
\mathcal{J}_{\mu}=\mathcal{R}^{2}\big(\nabla_{\mu}\mathcal{S}\big)~.
\end{equation}
The last term in the first relation above is the quantum potential, which we shall denote as  $f(\mathcal{R})$, 
defined by (with $\hbar^{2}$ restored)  $f({\mathcal R})=\frac{\hbar^{2}\Box\mathcal{R}}{\mathcal{R}}$. 
After one defines the appropriate curved space generalization of the four-momentum of the 
particle  i.e., $p_{\mu}=\nabla_{\mu}\mathcal{S}$, this equation gives the constraint on the  magnitude of four-momentum
\begin{equation}\label{const}
p_{\mu} p^{\mu}=-m^{2}-\epsilon R+\frac{\Box\mathcal{R}}{\mathcal{R}}~.
\end{equation}
On the other hand the second relation above represent a conservation equation for the current $\mathcal{J}^{\mu}$. 
	
From the above eq.(\ref{const}), we glean that  the norm of the four-momentum is not constant, and given this 
magnitude of the four-momentum it will be important how one defines a four-velocity vector from it. 
We do this  by defining the normalized four-velocity for a timelike trajectory as $u^{\mu}u_{\mu}= -1$, so that \cite{SG}
\begin{equation}\label{mass}
p^{\mu}=M(x)u^{\mu},~~ M(x)=\sqrt{\big(m^{2}-f(\mathcal{R})+\epsilon R\big)}~.
\end{equation}
With this definition, the four-velocity remains the same as in the classical trajectory but the particle's 
mass becomes a variable depending on the quantum amplitude $\mathcal{R}$, and this is known as the 
quantum mass of the particle. It is known that under an appropriate conformal transformation the action
of a particle of variable mass becomes the action of a particle of
constant mass (and vice versa). Using this fact in the next section we
shall work in a conformally related frame, with conformal factor being
appropriate function of quantum potential such that the dispersion
relation of eq.(\ref{const}) transforms to the dispersion relation of a particle of
constant mass.

On the other hand the second  equation of eq.(\ref{kgri}) implies, with this interpretation of variable quantum mass, 
the conservation equation of the 4-current to be  $\nabla_{\mu}\big(\mathcal{R}^{2}M u^{\mu}\big)=0$. 
Unfortunately however, now it is not possible  to interpret this equation (as is done in the corresponding 
non-relativistic Bohmian treatment of Schrodinger equation) as a continuity equation with the probability 
density  defined as $\rho=\mathcal{R}^{2}$. Because the corresponding equation should look like 
$\nabla_{\mu}\big(\rho^{2} u^{\mu}\big)=0$ \cite{TT}. As mentioned above, one of the motivation to transform to conformal 
frame is to make the particle mass constant and hence the equation of motion a geodesics, but it is not 
possible (as we shall explain in subsection \ref{ceq} below), with the same transformation to make the 
conservation equation a continuity equation for probability density. Recently in \cite{JCa} 
the authors have suggested a solution of this problem. In this work we shall propose an alternative 
one -  use a disformal transformation rather than the conformal one.
	
As is clear from eq.(\ref{const}), due to presence of the quantum potential (the $f(\mathcal{R})$ term),
the equation of motion of the particle is not a geodesic. Rather it contains an extra term coming form the 
force generated by the quantum potential. The straightforward way to find this force term in the 
acceleration equation is to take the directional derivative of the constraint relation of eq.(\ref{const}) 
by introducing a parameter (say $\tau$) along the particle trajectory (see \cite{RG} for a derivation along these lines).
For later purposes however, we shall use a variational principle and  write down  the action along the particle
trajectory between two points (say $1$ and $2$) as
\begin{equation}\label{action}
S[x^{\mu}(\tau),\eta(\tau)]=\int_{1}^{2}d\tau \mathcal{L}, ~~~ \mathcal{L}=\frac{1}{2}\bigg[\eta^{-1}
g_{\mu\nu}u^{\mu}u^{\nu}-\eta\Big(m^{2}-f(\mathcal{R})+\epsilon R\Big)\bigg]~,
\end{equation}
where $\eta$ is a Lagrange multiplier, and $u^{\mu}$ is the four-velocity of the particle,
normalized as $u^{\mu}u_{\mu}=-1$.\footnote{This four-velocity normalization is different from 
\cite{SD} but same as \cite{RG}.} Variation of the  action with respect to $x^{\mu}$ gives the usual Euler-Lagrange equations,
and a variation with respect to the Lagrange multiplier gives $\eta^{2}=-\frac{u^{\mu}u_{\mu}}{M^{2}}$, 
with $M$  being the quantum mass defined above. Now substituting this into the Lagrangian of eq.(\ref{action}), we have  
\begin{equation}
\label{Lagrangian}
\mathcal{L}=\eta^{-1}\big[u^{\mu}u_{\mu}\big]=-M\sqrt{-g_{\mu\nu}u^{\mu}u^{\nu}}~,
\end{equation} so that  the 
corresponding four-momentum $p_{\mu}=\frac{\partial \mathcal{L}}{\partial \dot{x}^{\mu}}=\frac{M}{\sqrt{-u^{\mu}u_{\mu}}}u_{\mu}$
satisfies the required constraint relation $p_{\mu}p^{\mu}=-M^2$ (irrespective of the normalization of the 4-velocity).
The acceleration equation corresponding to this Lagrangian is obtained by using the  Euler-Lagrange equations, and is given by 
\begin{equation}\label{geo1}
u^{\mu}\nabla_{\mu}u^{\nu}=-\frac{1}{2}\big(g^{\mu\nu}+u^{\mu}u^{\nu}\big)\nabla_{\mu}\ln(M^{2})=-\frac{1}
{2}\big(g^{\mu\nu}+u^{\mu}u^{\nu}\big)\nabla_{\mu}\ln\Big[1-\Big(f(\mathcal{R})-\epsilon R\Big)/m^{2}\Big]~.
\end{equation}
As anticipated, the right hand side of eq (\ref{geo1}) is non zero, and this indicates that the motion of a particle 
which is freely falling in the classical description is no longer force free for motion along a Bohmian trajectory. 
This term comes from the quantum potential. So unless $\mathcal{R}=constant$, or it satisfies the equation $\Box\mathcal{R}=0$, this term is non-zero. This matches with the acceleration equation derived in \cite{RG} 
from the momentum constrain relation (\ref{const}), but it is different from the equation written in \cite{SD}. 
This is due to the different parameterization employed, namely, the term proportional to $u^{\mu}u^{\nu}$ can be absorbed 
by changing the parameter, although the first term can never be removed by any such redefinition of the parameter.  
Also note  that the force term (from eq.(\ref{geo1})) is perpendicular to the velocity vector, i.e.,
$u^{\nu} u^{\mu}\nabla_{\mu}u_{\nu}=u^{\mu}h_{\mu\nu}=0$, 
where we have defined the transverse metric $h_{\mu\nu}=g_{\mu\nu}+u_{\mu}u_{\nu}$.

Now consider a congruence of timelike particle trajectories with $u^{\mu}$ being the tangent to the trajectories. In a standard
fashion \cite{EP}, we consider $B_{\mu\nu}=\nabla_{\nu}u_{\mu}$ and calculate it's change along the particle trajectories as
\begin{equation}
u^{\mu}\nabla_{\mu}B_{\alpha\beta}=\nabla_{\beta}(u^{\mu}\nabla_{\mu}u_{\alpha})-B_{\alpha\mu}B^{\mu}_{\beta}-
R_{\alpha\mu\beta\nu}u^{\mu}u^{\nu}.
\end{equation}
Taking the trace of this equation, and using the acceleration equation derived above, we arrive at the 
modified Raychaudhuri equation given by
\begin{equation}\label{quRay}
\frac{d\theta}{d\tau}=h^{\mu\nu}\Big(\nabla_{\mu}\nabla_{\nu}L+\nabla_{\mu}L\nabla_{\nu}L\Big)+
\theta \frac{dL}{d\tau}-\frac{1}{3}\theta^{2}-\sigma^{\alpha\beta}\sigma_{\alpha\beta}+\omega^{\alpha\beta}
\omega_{\alpha\beta}-R_{\alpha\beta}u^{\alpha}u^{\beta}~,
\end{equation}
where $a^{\alpha}=u^{\mu}\nabla_{\mu}u^{\alpha}$ is the acceleration vector, $\theta=B^{\mu}_{\mu}=\nabla_{\mu}u^{\mu}$ 
is the expansion scalar, $\sigma_{\mu\nu}=B_{(\mu\nu)}-\frac{1}{3}\theta h_{\mu\nu}$ is the shear tensor and 
$\omega_{\mu\nu}=B_{[\mu\nu]}$ is the rotation tensor, and we have defined the function
\begin{equation}\label{cf}
L=-\frac{1}{2}\ln\Big[1-\Big(f(\mathcal{R})-\epsilon R\Big)/m^{2}\Big]~.
\end{equation} 
	
The terms involving $L$ in the quantum Raychaudhuri equation are the contributions of the 
quantum effects, and the relative contribution of these terms with respect to other classical terms  
determine whether the trajectory will reach a conjugate  point or not \cite{SD}.
Note that this form of Raychaudhuri equation is different from one given in \cite{SD}, 
due to the different parameterization of the trajectory. It is also different from the one derived in 
\cite{RG}, because those authors took $\nabla_{\alpha} h_{\mu\nu}=0$.
We also record the expression of the deviation equation for the vector field $\xi^{\mu}$ between two non-geodesic curves as  
\begin{equation}\label{qude}
\frac{d^{2}\xi^{\mu}}{d\tau^{2}}=\xi^{\alpha}\Big(u^{\mu}u^{\nu}\nabla_{\alpha}\nabla_{\nu}L+u^{(\mu}
\nabla_{|\alpha|}u^{\nu)}\nabla_{\nu}L\Big)+R^{\mu}_{\alpha\rho\nu}\xi^{\alpha}u^{\rho}u^{\nu}.
\end{equation}

\subsection{Bohmian trajectories in a conformally related frame}

The purpose of this subsection is to discuss the transformation of the acceleration equation derived above,
under a conformal transformation. If two metrics are related by a conformal transformation of the form 
$\tilde{g}_{\mu\nu}=\Omega^{2}(x)g_{\mu\nu}$, 
then it is well known that the relation between the acceleration equations in two frames are given by (see \cite{IQ})
\begin{equation}\label{conformalacc}
\tilde{u^{\mu}}\tilde{\nabla}_{\mu}\tilde{u}^{\nu}= u^{\mu}\nabla_{\mu}u^{\nu}+h^{\mu\nu}\nabla_{\mu}(\ln \Omega)~.
\end{equation}
From this relation, we see that if a particle moves along the geodesic of the transformed frame,
then the motion in the frame $g_{\mu\nu}$ is non geodesic.

In our case, we choose to work in a frame that is  related to the original metric by a conformal factor, such that the relation 
between the two metrics are\footnote{The conformal factor $\Omega^{2}$ is a dimensionless quantity. But in the following
we shall ignore the constant $m^{2}$ factor.}
\begin{equation}
\tilde{g}_{\mu\nu}=(M^{2}/m^{2}) g_{\mu\nu}~.
\label{conformal}
\end{equation}
We can conventionally express the above relations in terms of the action of a particle corresponding to the 
Lagrangian of eq.(\ref{Lagrangian}), along the path $\gamma$ given by $x^{\mu}=x^{\mu}(\tau)$,
\begin{equation}\label{act}
S_{m}\big[g_{\mu\nu},\gamma,M\big]=-\int_{\gamma}m\sqrt{1-\frac{f(\mathcal{R})-\epsilon R}{m^{2}}}
\sqrt{-g_{\mu\nu}u^{\mu}u^{\nu}}d\tau,~~u^{\mu}=\frac{dx^{\mu}}{d\tau}.
\end{equation}
In the conformally transformed frame the corresponding action for a path $x^{\mu}=x^{\mu}(\tilde{\tau})$ is transformed into 
\begin{equation}\label{acc}
S_{m}\big[\tilde{g}_{\mu\nu},\gamma\big]=-\int_{\gamma}\sqrt{-\tilde{g}_{\mu\nu}\tilde{u}^{\mu}\tilde{u}^{\nu}}d\tilde{\tau}~,~~
u^{\mu}=\frac{dx^{\mu}}{d\tilde{\tau}}.
\end{equation}
As can be readily seen by comparing them, the first form represents a particle having a variable mass $M$ defined in 
eq.(\ref{mass}), and the second one is the action for a unit mass particle. If we write down the acceleration equations 
corresponding to these actions, they turn out just to be same equations derived   before i.e.,
 $ u^{\mu}\nabla_{\mu}u^{\nu}+h^{\mu\nu}\nabla_{\mu}(\ln \Omega)$ and $\tilde{u^{\mu}}\tilde{\nabla}_{\mu}\tilde{u}^{\nu}=0$, and thus satisfy eq.(\ref{conformalacc}). 
The implication of this relation is straightforward. Namely, the motion in the conformal frame is free from the force of the 
quantum potential. If we denote the momentum corresponding to the conformal frame as $\tilde{p}^{\mu}$, then this 
can be shown to satisfy $\tilde{p}^{\mu}\tilde{p}_{\mu}=-1$. 
	
If the motion in the conformal frame is force free, then where are the effects of the quantum potential in such a frame? 
The answer is simply that it is hidden inside the metric $\tilde{g}_{\mu\nu}$ via the conformal factor. In other words 
the quantum effects are in the field equations, so that the metric and also the energy momentum tensors are different.
Indeed, starting from the action of eq.(\ref{EHA}), making the  transformation of eq.(\ref{conformal}), 
we obtain, after some algebra the following simplified action
\begin{equation}\label{Coac}
S=\int d^{4}x\sqrt{-\tilde{g}}\tilde{\Phi}\Big[\epsilon\tilde{R}\tilde{\Phi}-\tilde{\Box}\tilde{\Phi}
+M^{-2}m^{2}\tilde{\Phi}\Big] +\int d^{4}x\sqrt{-\tilde{g}}\tilde{\mathcal{L}_{m}}~,
\end{equation}
where we have redefined $\tilde{\Phi}=M^{-1}\Phi$. Since the conformal factor $M^{2}$ is a real (or purely imaginary) number,  
$\tilde{\Phi}$ and $\Phi$ have different norms ($\mathcal{R}$), but they correspond to  the same four-momentum 
$p_{\mu}=\nabla_{\mu}\mathcal{S}$.
  
The standard variation of this action with respect to the modified 
metric $\tilde{g}_{\mu\nu}$ and the modified scalar field $\tilde{\Phi}$ gives the  conformal version of 
Einstein equations and the scalar field equation (which are standard, see, e.g., \cite{carroll} and \cite{dab}). 
The energy momentum tensors in the two frames are related by $\tilde{T}^m_{\mu\nu} = M^{-2}T^m_{\mu\nu}$
and hence transformed frame the EM tensor is no longer conserved, unless the trace of the EM tensor $\tilde{T}_{m}$
in the conformal frame vanishes, i.e., $\tilde{\nabla}_{\nu}\tilde{T}_{m}^{\mu\nu}=-\tilde{T}_{m}(\nabla^{\mu}\ln M)$.

\subsection{Conserved quantities }\label{Ckv}
	
If one wants to study the observational aspect of the problem of a  particle moving along a quantum trajectory (eq.(\ref{geo1})), 
in presence of a massive gravitational object (say a black hole), it is important to find out the 
conserved quantities. Because the motion is not that of a freely falling particle, it is not obvious if usual 
conserved quantities for stationary spacetime, namely energy and angular momentum are also conserved 
along Bohmian trajectories. 

When a classical particle of mass $m$ moves in a geodesic satisfying $p^{\mu}\nabla_{\mu}p_{\nu}=0$ 
(here $p^{\mu}=mu^{\mu}$), the quantity $K_{\nu}p^{\nu}$ is conserved along the motion of the particle,
i.e., $p^{\mu}\nabla_{\mu}(K_{\nu}p^{\nu})=0$, where $K^{\mu}$ is a Killing vector which satisfies the Killing 
equation $\nabla_{(\mu}K_{\nu)}=0$. 
To find out how this equation is modified when the motion is along a quantum Bohmian trajectory,
we take  covariant derivative of the constraint relation $p_{\mu}p^{\mu}=-M^{2}$ to obtain $p^{\mu}\nabla_{\mu}p_{\nu}=
-M\nabla_{\nu}M(x)$. Now along quantum trajectory, we find  
\begin{equation}\label{Kie}
p^{\mu}\nabla_{\mu}(K_{\nu}p^{\nu})= p^{\mu}K_{\nu}\nabla_{\mu}(p^{\nu})+p^{\mu}p^{\nu}\nabla_{\mu}K_{\nu}\\
=-K^{\nu}\big[M\nabla_{\nu}M(x)\big]+p^{\mu}p^{\nu}\nabla_{(\mu}K_{\nu)}~.
\end{equation}
We see that the Killing equation  no longer  implies that $K_{\nu}p^{\nu}$ is a conserved quantity.	
However, it can be checked that $p_{\mu}K^{\mu}$ is a conserved quantity, if $K^{\mu}$ is a conformal Killing vector 
associated with the metric $\tilde{g}_{\mu\nu}$, i.e., it satisfies the equation 
$\nabla_{(\mu}K_{\nu)}=-g_{\mu\nu}K^{\lambda}(\nabla_{\lambda}\ln \Omega)$ (here $\Omega=M$).
	
The above conclusion is true for any general spacetime. For the special case of stationary spacetimes however, we 
can have an interesting situation, namely that, even for an ordinary Killing vector of  $g_{\mu\nu}$,
we can find a conserved quantity. To see this clearly, we start from  eq.(\ref{Kie}) and after a bit of algebra we arrive at
\begin{equation}\label{Ki3}
p^{\mu}\nabla_{\mu}(K_{\nu}p^{\nu})
=K_{\nu}Mu^{\mu}u^{\nu}\nabla_{\mu}M-K_{\nu}M^{2}h^{\mu\nu}\nabla_{\mu}
\ln M+M^{2}u^{\mu}u^{\nu}\nabla_{\mu}K_{\nu}~~.
\end{equation}
Now suppose that in the particle's wave function, $\mathcal{R}(x^{\mu})$ is independent of time so that 
the quantum potential term and hence $M$ are also time independent. In this case the vectors $\nabla_{\mu} M$ 
and $\nabla_{\mu}\log M$ have vanishing components along the $dt$ direction. If the background spacetime is 
stationary, and the velocity is timelike, i.e., $u^{\mu}=(1,0,0,0)$, then for a vector filed of the form 
$K^{\mu}=(1,0,0,0)$, the first two terms of eq.(\ref{Ki3}) are identically zero (note that $h^{\mu\nu}$ is spacelike in this 
case). This means that for a stationary spacetime and time independent  $\mathcal{R}$, 
once again  $K_{\mu}p^{\mu}$ is conserved along the trajectory when the Killing 
equation is satisfied. This fact can be used to generate a  quantum corrected version of any stationary spacetime 
(In \cite{AKh} such corrections to Schwarzschild solution is obtained by using the quantum Raychaudhuri equation).
	
As an immediate application of this, we can use the  conformal transformation 
above find out a conformally transformed  version of the Schwarzschild metric. Let us   
assume a simple stationary state  wave function so that that its modulus is given by 
$\mathcal{R}(r)=Nr\exp\Big({-\frac{r^{2}}{2}}\Big)$, where $N$ is a normalization constant. 
Then we calculate the quantum potential associated with the wave function on Schwarzschild background 
\begin{equation}
f(\mathcal{R}(r))=\frac{\Box\mathcal{R}}{\mathcal{R}}=\frac{1}{r^{3}}\Big[r\Big(r^{4}-5r^{2}+2\Big)-
2\mathcal{M}\Big(r^{4}-4r^{2}+1\Big)\Big]~,
\end{equation}
$\mathcal{M}$ being the Schwarzschild mass. The conformal version of Schwarzschild solution is given by,
\begin{equation}\label{qmSch}
d\tilde{s}^{2}=\frac{1}{r^{3}}\Big[2\mathcal{M}\big(r^{4}-4r^{2}+1\big)-r\big(r^{4}+2\big)\Big]\Big\{\Big(1-\frac{2\mathcal{M}}{r}\Big)~, 
\Big(1-\frac{2\mathcal{M}}{r}\Big)^{-1},r^{2}d\Omega^{2}\Big\}~,
\end{equation}
where we have neglected a constant additive term in the conformal factor. 
This is the solution of the transformed field equation  with $\tilde{T}^{m}_{\mu\nu}=0$.
A particle will follow the geodesics of this metric. The matter part of EM tensor is zero due to the 
relation $\tilde{T}^m_{\mu\nu} = M^{-2}T^m_{\mu\nu}$ discussed above. 
	
\subsection{Singularities in the conformal frame}
	
It was argued in \cite{SD} that since Bohmian trajectories cannot intersect each other, they do not form conjugate 
points and hence can avoid the usual singularities of GR. In this context, we ask the following question. 
In the conformally transformed frame there is no quantum force on the trajectory, then what happens to the 
conjugate points? Can the trajectories reach the  singularity in the conformal frame $\tilde{g}_{\mu\nu}$?

To answer this question, we consider the following conjecture proposed long back in \cite{Ke}. If the metric $g_{\mu\nu}$ is singular, 
we can always go to the conformal frame $\tilde{g}_{\mu\nu}=\Omega^{2}g_{\mu\nu}$ which is non-singular and  
the singularities of the original metric show up as zeros of the conformal factor (as usual, the zeros of the conformal factor 
represent the boundaries of the spacetime). Now the question reduces to whether the conformal 
transformation of eq.(\ref{conformal}) is the required one to ensure the nonsingular nature of $\tilde{g}_{\mu\nu}$. 
If we consider a particle of mass $m$ moving in a geodesics of the metric $\tilde{g}_{\mu\nu}$, then in the frame  $g_{\mu\nu}$, the particle's mass is time and position dependent (the modified mass 
in general depends on the nature of the conformal factor $\Omega$) and it moves on a nongeodesic. 
In \cite{Ke}, the author showed working out  several examples, that the variable mass of the conformal frame 
should be the conformal factor to ensure the nonsingular nature of the metric $\tilde{g}_{\mu\nu}$. From eq.(\ref{conformal}), 
we see this is exactly the case. 
	
This should be clear from the example we have worked out above. The original Schwarzschild metric is 
singular at $r\rightarrow 0$. But the Ricci curvature scalar  $\tilde{R}$ of the metric $\tilde{g}_{\mu\nu}$ is non-singular 
in this limit, as can be checked explicitly, i.e.,  $\tilde{R}_{r\rightarrow 0}= constant$. Also the conformal factor 
in the transformation equation in eq.(\ref{qmSch}) is undefined in this limit. Thus if we start from the nonsingular 
frame $\tilde{g}_{\mu\nu}$ and work in a conformally related frame $g_{\mu\nu}$, the conformal 
factor vanishes precisely at $r=0$ making the transformation invalid at the singularity $r=0$.  
Thus we conclude that the singularity is resolved in both the frames. Note that the quantum potential and 
hence the conformal factor depends on the wave function, and thus it is not guaranteed that the transformed metric 
is generically singularity free. This general case will be investigated elsewhere. 
	
\subsection{Problems with a conformally related frame}\label{ceq}

So far we have shown by the transformation of eq.(\ref{conformal}), that 
it is possible to make the magnitude of $\tilde{p}_{\mu}$ a constant. By using the same transformation, can 
we write the continuity equation (second equation in eq.(\ref{kgri})) in the desired form 
$\tilde{\nabla}_{\mu}\big(\rho\tilde{u}^{\mu}\big)=0$? The answer is no. To explain this, we shall first derive an 
important relation between the expansion scalars ($\theta$ and $\tilde{\theta}$) of two frames
and the probability density $\rho$ (see eq.(\ref{cthr}) below).

A conformal transformation is equivalent to a corresponding change in the proper time 
$d\tau \rightarrow d\tilde{\tau}=\Omega d\tau$. In GR, under such conformal transformations, 
the Christoffel symbols transform as \cite{carroll, RW}
\begin{equation}\label{Chtr}
\delta\Gamma^{\sigma}_{\mu\nu}=2\delta^{\sigma}_{(\mu}\partial_{\nu)}\ln \Omega-g_{\mu\nu}\partial^{\sigma}\ln \Omega~.
\end{equation}
Using this, one can check that the covariant derivative of the four velocity in the transformed frame is
\begin{equation}\label{4tr}
\tilde{\nabla}_{\mu}\tilde{u}_{\nu}=\Omega\Big(\nabla_{\mu}u_{\nu}+g_{\mu\nu}u^{\alpha}
\nabla_{\alpha}\ln \Omega-u_{\mu}\nabla_{\nu}\ln \Omega\Big)~. 
\end{equation}
Taking the trace of this equation, we see that the expansion scalar $\theta$ of a geodesic congruence 
transforms under a conformal transformation as
\begin{equation}\label{Cth}
\tilde{\theta}=\Omega^{-1}\Big(\theta+3u^{\alpha}\nabla_{\alpha}\ln\Omega\Big)~.
\end{equation}
Using this relation we derive the following relation involving the probability density $\rho$,
\begin{equation}\label{cthr}
\tilde{\nabla}_{\mu}\big(\rho\tilde{u}^{\mu}\big)=\rho\tilde{\theta}+\tilde{u}^{\mu}\tilde{\nabla}_{\mu}\rho=
\rho\Omega^{-1}\Big(\theta+3u^{\alpha}\nabla_{\alpha}\ln\Omega\Big)+\Omega^{-1}u^{\mu}\nabla_{\mu}\rho~.
\end{equation}
The criterion that in the transformed frame, the particle motion is a geodesic has already fixed the 
conformal factor $\Omega$ equal to $M$ up to a multiplicative constant. Now expanding  the original 
conservation equation $\nabla_{\mu}\big(\rho M u^{\mu}\big)=0$ we have the relation
\begin{equation}\label{rhth}
\rho\theta=-\rho u^{\alpha}\nabla_{\alpha}\ln M- u^{\alpha}\nabla_{\alpha}\rho~.
\end{equation}
Eliminating $\theta$ between eqs.(\ref{cthr}) and (\ref{rhth}), we see that $\tilde{\nabla}_{\mu}\big(\rho\tilde{u}^{\mu}\big)\neq0$. 
Thus transformation to a conformal frame gives rise to a wrong continuity equation. The reason for this is of course the fact that 
the momentum constrain relation has already fixed the conformal factor equal to $M$. As we shall show in sequel, for disformally related metrics,
there is still enough freedom to fix both the problems. 
	
A somewhat subtle point here is worth mentioning. Remember that we redefined the scalar field as 
$\Phi=M\tilde{\Phi}$, so that the resulting action in eq.(\ref{Coac}) is more convenient to work with. 
As we have mentioned, in this redefinition, the norm of the wave function changes to $\tilde{\mathcal{R}}=M^{-1}\mathcal{R}$, 
and hence if one defines a transformed probability density $\tilde{\rho}=\tilde{\mathcal{R}}^{2}$ and demands 
that $\tilde{\nabla}_{\mu}\big(\tilde{\rho}\tilde{u}^{\mu}\big)$ is the quantity one should be looking for, we can see, by an 
analogous procedure as above that this quantity is indeed conserved, i.e., 
$\tilde{\nabla}_{\mu}\big(\tilde{\rho}\tilde{u}^{\mu}\big)=0$. But let us stress that this is not the continuity 
equation we are after, simply because the field redefinition (Weyl scaling) has nothing to do  
with the original conformal transformation $g_{\mu\nu}\rightarrow\tilde{g}_{\mu\nu}$, which is an actual change of the 
geometry itself, and not a change of coordinates or fields living in the spacetime \cite{RW}. 
The only purpose it serves (in this context) is to rewrite the action in a convenient form. Notice also that  
when $M^{2}<0$ the transformed density ($\tilde{\rho}$) is not even well defined in the transformed frame. 
	
Apart from the wrong continuity equation, there is a further issue that is problematic in a conformally
related frame. So far in our discussion of non geodesic motion, we have always consider a timelike 
trajectory for which $u^{\mu}u_{\mu}=-1$, but we can generalize the result for spacelike and null trajectories also. 
For the general case the eq (\ref{geo1}) is given by
\begin{equation}\label{geo11}
u^{\mu}\nabla_{\mu}u^{\nu}=\frac{1}{2}\big(\alpha g^{\mu\nu}- u^{\mu}u^{\nu}\big)\nabla_{\mu}\ln\big(m^{2}-
f(\mathcal{R})\big)~, 
\end{equation}
where $\alpha=-1,+1,0$, for time-like, space-like and null geodesics, respectively. 
Now for a massless particle, we put $m=0$, and since the term proportional to $u^{\mu}u^{\nu}$ can be absorbed  
in a re-parameterization of proper time, the only remaining term is proportional to $\alpha$, and hence the 
acceleration is zero for massless particle in both frames. So a natural question is, 
if null trajectories always move in geodesics, what happens to the singularity resolution argument for 
massless particles ? That is, if the force due to the quantum potential does not affect null trajectories,  
we ask why they do not form a caustic as in GR, and hence fall into a singularity?
	
We will next show that all the problems mentioned above are resolved by using disformal transformations instead. However
before going into this, we will make an important simplification by assuming minimal coupling. 

\subsection{Description in a  minimally coupled background}\label{minc}

The analysis of the quantum corrections so far has been carried out in a background where there is a 
non-minimal coupling between the curvature scalar $R$ and the scalar field describing the particle. 
As mentioned this term is crucial for the Klein-Gordon equation to be conformally invariant. 
Though this helps to make the analysis in the conformal frame simpler, it makes a nontrivial contribution 
to the Einstein equation, such that the metric in general will not be non-dynamical - as can be seen from eq.(\ref{EEJF}). 
Usually this contribution is neglected (as was done in \cite{SD}) and the scalar field is assumed to be defined 
on a non-dynamical static background. 
	
From now on we will instead work in a frame where $\Phi$ is minimally coupled to the background. 
Then the total action is similar to the one in
eq.(\ref{EHA}) with $F(\Phi)=1$, and $\Phi$ satisfies the Klein-Gordon equation $(\Box -m^{2}) \Phi =0$.
The dispersion relation now changes to
\begin{equation}
p_{\mu} p^{\mu}=-m^{2}+\frac{\Box\mathcal{R}}{\mathcal{R}}=M^{2}u_{\mu}u^{\mu}~,
\label{dispersion}
\end{equation}
which indicates that a quantum particle will follow a non geodesics motion in this frame as given in eq.(\ref{geo1}),
with $\epsilon=0$. We can always transform to a conformal frame, where the motion of the particle will be on  
a geodesic and quantum effects are included in the energy momentum tensor, but in that frame there is a 
non-minimal coupling between the curvature scalar and the redefined scalar field\footnote{For a massless particle 
one can make the action and hence the Klein-Gordon equation conformally invariant by choosing $\epsilon=1/6$ in 4d. But for
massive particles, with which we shall mostly work with, this invariance is lost.}.
	
The advantages of this approach to that the picture is more in line with scalar tensor theories of gravity, where in the non-minimally 
coupled frame (Jordan frame) the particle follows geodesic equation, but in the conformally related frame,  
where $\Phi$ is minimally coupled, the particle follows a non geodesic motion.  
Also, in this frame, the quantum mass, given by $M=\sqrt{m^{2}-f(\mathcal{R})}$ is independent of the nature of the 
coupling term $\epsilon$, and hence its value is unambiguous i.e., if $f(\mathcal{R})>m^{2}$  it is imaginary, 
otherwise it is real. In the Jordan frame, this depends on $\epsilon$. As we will momentarily see, this property  
helps to solve the problem of indefiniteness of quantum mass uniquely in 
this frame, using a disformal transformation.  

All the calculations in the rest of the paper are performed assuming a minimal coupling.  
When there is non-minimal coupling between gravitational and particle degrees of freedom, our calculations 
can be generalized with some minor modifications, but in the transformed frame it is rather difficult to see the 
quantum effects in the metric through calculating the EM tensor, because there will be terms coming from the 
non-minimal coupling also (with minimal coupling, the last two terms in the right hand side of eq.(\ref{EEJF}) are zero). 
In short, when there is already a coupling between a classical background and the  degrees of freedom of the 
quantum particle, the process of incorporating quantum effects in background geometry loses some of its 
meaning because from the start the classical and quantum degrees of freedom are related.

\section{Bohmian trajectory on a physical geometry}\label{sec3}
	
So far we have assumed a  description of gravity where gravitational and particle dynamics take place in conformally 
related global Riemannian spacetimes, called gravitational and physical geometries respectively. 
But as we have pointed out in the introduction, the relation between these two geometries can be more general 
than the conformal transformation. It was shown in \cite{JBe}, by assuming the physical geometry 
(on which the matter dynamics takes place) to be Finslerian (instead of Riemannian), and using arguments 
based on the weak equivalence principle and causality,  that in the most general case, both the gravitational and 
the physical geometries have to be Riemannian and that in general such geometries 
are related to each other by a disformal transformation. It is thus natural to ask what are the consequences of
quantum motion in terms of Bohmian mechanics in this more general case.  This is the topic we discuss in 
this section. 
	
\subsection{Disformal transformations}

Two metrics $g^{*}$ and $g$ (and their inverses) are said to be related to each other by the a disformal transformation, 
if the relation between them is given by \cite{JBe,CCY}
\begin{equation}\label{Dist}
g^{*}_{\mu\nu}=\Omega^{2}(\phi,X)g_{\mu\nu}-\alpha \mathcal{B}(\phi, X)\phi_{\mu}\phi_{\nu}~, ~~ g^{*\mu\nu}
=\Omega^{-2}\bigg[g^{\mu\nu}+\alpha \frac{\mathcal{B}}{\Omega^{2}-2X\mathcal{B}}\phi^{\mu}\phi^{\nu}\bigg]~, 
~~ X=-\frac{1}{2}g^{\mu\nu}\phi_{\mu}\phi_{\nu}~,
\end{equation}
where $\alpha = 0, \pm 1$, both $\Omega$ and $\mathcal{B}$ are arbitrary real functions of a scalar field $\phi$, and we have used 
the notation  $\phi_{\mu}=\nabla_{\mu}\phi$ to denote the normal vector to a $\phi=const$ hypersurface. 
Note that all indices of $\phi_{\mu}$ are raised and lowered by the metric $g_{\mu\nu}$. 
We will denote all the tensor quantities with respect to $g^{*}_{\mu\nu}$ with a superscript ``$*$''.

If we consider the motion of a particle moving on a timelike trajectory we can choose  the normal 
vectors $\phi_{\mu}$ to be  hypersurface orthogonal velocity vector $v^{\mu}$ of the trajectories, and in this case 
$\alpha=-1$ (though in the calculations below we will keep $\alpha$ to generalize the  results to a spacelike trajectories also). 
The identification of $\phi_{\mu}$ with velocity  is a choice, and as long as we are considering the motion 
along timelike (or spacelike) trajectory this is a good choice, but for null trajectories we will make a  different 
choice because in general via a disformal transformation, unlike a conformal one, a null geodesic 
can map to a non-geodesic trajectory (this will be crucial in our arguments in section \ref{m0}).  Thus the form of the 
disformal transformation and its inverse that we will consider for now are the followings \cite{DK}
\begin{equation}\label{sdt}
g^{*}_{\mu\nu}=\Omega^{2}(x)g_{\mu\nu}-\alpha \mathcal{B}(x)v_{\mu}v_{\nu}~, ~~ 
g^{*\mu\nu}=\Omega^{-2}\bigg[g^{\mu\nu}+\alpha \bigg(\frac{\mathcal{B}}{\Omega^{2}-\mathcal{B}}\bigg)v^{\mu}v^{\nu}\bigg]~.
\end{equation}
This transformation is equivalent to a transformation of the  proper time \cite{CCY}
\begin{equation}
d\tau^{*2}=-g^{*}_{\mu\nu}dx^{\mu}dx^{\nu}=-\Big[\Omega^{2}g_{\mu\nu}-\alpha
\mathcal{B}v_{\mu}v_{\nu}\Big]dx^{\mu}dx^{\nu}=\Big[\Omega^{2}+\alpha \mathcal{B}\Big]d\tau^{2}~.
\end{equation}
Now, the tangent vector to a particle trajectory is $v^{\mu}=dx^{\mu}/d\tau$. With respect to 
$g_{\mu\nu}^{*}$, this vector is defined as $v^{*\mu}=dx^{\mu}/d\tau^{*}$. From the above relations we can find out the following relations
\begin{equation}
v^{*}_{\mu}=\sqrt{\mathcal{F}}v_{\mu}~, ~~ v^{*\mu}=g^{*\mu\nu}v^{*}_{\nu}=\mathcal{F}^{-1/2}v^{\mu}~,~~
\mathcal{F}=\Omega^{2}-\mathcal{B}, ~~ g^{*}_{\mu\nu}v^{*\mu}v^{*\nu}=\alpha.
\end{equation}
As can be seen, the vector $v^{*\mu}$ is normalized with respect to $g^{*\mu\nu}$.
	
\subsection{Relation between acceleration vectors and particle motion}\label{disacc}

Using the formulas given above, we can now establish the required  relation between  acceleration 
vectors $a^{\nu}=v^{\mu}\nabla_{\mu}v^{\nu}$ and $a^{*\nu}=v^{*\mu}\nabla^{*}_{\mu}v^{*\nu}$  of two 
metrics, where $\nabla^{*}_{\mu}$ is the covariant derivative with respect to $g^{*}_{\mu\nu}$. 
First, we write down the known relation between the quantities $\Gamma^ {*\lambda}_{\mu\nu}v^{*}_{\lambda}$ 
and $\Gamma^ {\lambda}_{\mu\nu}v_{\lambda}$   (see \cite{DK} for details)
\begin{equation}\label{gamma}
\Gamma^ {*\lambda}_{\mu\nu}v^{*}_{\lambda}=\sqrt{\mathcal{F}}\Gamma^ {\lambda}_{\mu\nu}v_{\lambda}
+\frac{1}{2\sqrt{\mathcal{F}}}\Big[\alpha v_{\mu}v_{\nu}v^{\alpha}\nabla_{\alpha}\mathcal{F}-
h_{\mu\nu}v^{\alpha}\nabla_{\alpha}\Omega^{2}-2\mathcal{B}K_{\mu\nu}+2v_{(\mu}h^{\alpha}_{\nu)}
\nabla_{\alpha}\mathcal{F}\Big]~,
\end{equation}
where $\Gamma^ {\lambda}_{\mu\nu}$ are the  connection coefficients, and we have defined the  extrinsic curvature 
in standard fashion, as
\begin{equation}
K_{\mu\nu}=h_{\mu}^{\alpha}\nabla_{\alpha}u_{\nu}=\nabla_{\mu}v_{\nu}-\alpha a_{\nu}v_{\mu}~.
\end{equation}
Substituting eq.(\ref{gamma}) in the formula 
$\nabla^{*}_{\mu}v^{*}_{\nu}=\partial_{\mu}v^{*}_{\nu}-\Gamma^ {*\lambda}_{\mu\nu}v^{*}_{\lambda}$,
after a few steps of straightforward algebra, we get the following relation between covariant derivative of a 
vector with respect to both metrics 
\begin{equation}\label{d*}
\nabla^{*}_{\mu}v^{*}_{\nu}=\sqrt{\mathcal{F}}\bigg(\nabla_{\mu}v_{\nu}+\frac{v^{\alpha}\nabla_{\alpha}\Omega^{2}}
{2\mathcal{F}}h_{\mu\nu}+\frac{\mathcal{B}}{\mathcal{F}}K_{(\mu\nu)}+\frac{v_{\nu}\nabla_{\mu}\mathcal{F}}
{2\mathcal{F}}-\frac{\alpha v_{\mu}v_{\nu}v^{\alpha}\nabla_{\alpha}\mathcal{F}}{2\mathcal{F}}-\frac{1}
{\mathcal{F}}v_{(\mu}h_{\nu)}^{\alpha}\nabla_{\alpha}\mathcal{F}\bigg)~.
\end{equation}
Note that the last three terms in eq.(\ref{d*}) are missing in \cite{DK}. But these terms are essential to 
establish a relation between the acceleration vectors, in which case both the second and third terms vanish. 
Multiplying both sides of eq.(\ref{d*}) by $v^{*\mu}$, and after a bit of  manipulation we get the following 
simplified version of the desired relation between the  acceleration vectors
\begin{equation}\label{Acc}
a^{*}_{\mu}=a_{\mu}-\frac{1}{2\mathcal{F}}\alpha h_{\mu}^{\nu}\nabla_{\nu}\mathcal{F}=a_{\mu}-\frac{1}{2}\alpha
h_{\mu}^{\nu}\nabla_{\nu}\ln\mathcal{F}~.
\end{equation} 
It is worth remarking that due to absence of the terms   in eq.(\ref{d*}) mentioned above in the corresponding formula in  
\cite{DK}, the author has  concluded that the accelerations are equal. 
But  accelerations of conformally (and also disformally) related frames are not equal, at least in GR. 
This is the root of the problem that one faces when constructing conformally invariant observables in 
gravity theories with symmetric connection (see the discussion is section \ref{sec5}  below). 

Note also that we shall recover all the 
results of previous section of conformal transformation by putting $\mathcal{B}=0$ in the above expressions.  
As is evident from this equation, two vectors  $a^{*}_{\mu}$ and $a_{\mu}$ cannot be equal to each other unless 
the difference of  $\Omega$ and $\mathcal{B}$ is a constant. This in turn means that two velocity vectors 
$v_{\mu}$ and $v^{*}_{\mu}$ cannot represent geodesics simultaneously. Indeed 
from eq.(\ref{Acc}), we glean that 
\begin{equation}\label{Diacc}
a^{*}_{\mu}=v^{*\nu}\nabla^{*}_{\nu}v^{*}_{\mu}=0 \implies ~~ a_{\mu}=v^{\nu}\nabla_{\nu}v_{\mu}=\alpha
h_{\mu}^{\nu}\nabla_{\nu}\ln\sqrt{\mathcal{F}}~.
\end{equation}
Thus in the physical geometry, if we consider the geodesic motion of a particle of mass $m$, then 
in the disformally related gravitational geometry, this represents an accelerated motion with the force 
being perpendicular to the four-velocity. 

Now if we consider a quantum particle moving along a Bohmian trajectory so that it is acted upon by the 
force whose origin is the quantum potential, then the norm of its  four momentum satisfies the relation 
in eq.(\ref{dispersion}). If we represent its acceleration by the  second equation in eq.(\ref{Diacc}), in  
the disformally related geometry this corresponds to force free motion with the quantum effects taken in 
the modified metric $g^{*}_{\mu\nu}$ of the physical geometry. We can easily determine the desired relation 
between the transformation factors $\Omega$ and $\mathcal{B}$ in 
terms of the modulus of the wave function $\mathcal{R}$, by comparing eq.(\ref{Diacc}) and  eq.(\ref{geo1}) 
to be\footnote{As before we will neglect the unimportant factor of $m^{2}$ and will take $\mathcal{F}=M^{2}$ below.},
\begin{equation}\label{omeB}
\mathcal{F}=\Omega^{2}-\mathcal{B}=\big(1-f(\mathcal{R})/m^{2}\big)~.
\end{equation}
We have put $\epsilon=0$ because we have assumed no coupling between the
curvature and the wavefunction $\Phi$. Note that the constraint relation of eq.(\ref{dispersion}) 
is not enough to uniquely determine the both the transformation factors in eq.(\ref{omeB}), and  below we will see that a 
second relation between the transformation factors arise through the continuity equation. 
	
It is also convenient to write down the corresponding particle actions in this case.
Given the constraint relation of eq.(\ref{dispersion}), by following the procedure outlined previously,
we can derive the action of a particle moving along the Bohmian trajectory in the frame $g_{\mu\nu}$. 
This is just given by eq.(\ref{act}) with $\epsilon=0$. Now using the rules of active transformation i.e.,
substituting $g_{\mu\nu}$ in terms of $g^{*}_{\mu\nu}$ and $v^{*\mu}$ from eq.(\ref{sdt}) we see that the action 
\begin{equation}
S_{m}\big[g_{\mu\nu},\gamma\big]=-\int_{\gamma}\sqrt{m^{2}-f(\mathcal{R})}\sqrt{-g_{\mu\nu}v^{\mu}v^{\nu}}d\tau
\end{equation}
gets transformed to the action of a particle of unit mass
\begin{equation}
S^{D}_{m}\big[g^{*}_{\mu\nu},\gamma\big]=-\int_{\gamma}\sqrt{-\Omega^{-2}\Big[g^{*}_{\mu\nu}
+\frac{\alpha\mathcal{B}}{\mathcal{F}}v^{*}_{\mu}v^{*}_{\nu}\Big]\mathcal{F}v^{*\mu}v^{*\nu}}d\tau^{*}
=-\int_{\gamma}\sqrt{-g^{*}_{\mu\nu}v^{*\mu}v^{*\nu}}d\tau^{*}~.
\end{equation}
Here in the first step, we have used the previous identification of eq.(\ref{omeB}), and also 
substituted $d\tau$ in terms of $d\tau^{*}$, and the trick in the second step is to write one 
set of $g^{*}_{\mu\nu}v^{*\mu}v^{*\nu}$ as $\alpha$ so that it cancels with the other term. 
With these particle actions, it is easy to check that they satisfy the respective equations 
in eq.(\ref{Diacc}), thus confirming our conclusions. Note that here (and in the previous section during the 
discussions on the conformal version) we have taken an active point of view in transforming the action, 
namely we have taken the functional form of the action same as before (i.e., the action of eq.(\ref{action})) with the 
Lagrangian in eq.(\ref{Lagrangian})), and replaced the original metric by the transformed one. This procedure 
naturally leads to a value of the action different from the previous one. On the other hand, if we want to keep the form  
of the action unchanged then the passive transformation should be used. In this paper we shall always use 
the active point of view unless otherwise specified (see \cite{dab,DS} for transformation with passive point of view).
	
We can also write down the modified Raychaudhuri equation and the deviation equation in this case also 
following the lines described in subsection \ref{kgequation} for the conformal frame. One can check  these are still 
given by  eq.(\ref{quRay}) and eq.(\ref{qude}) respectively, with  $L= -\frac{1}{2}\ln \mathcal{F}$.
	
\section{Advantages of disformal transformations}\label{sec4}
	
Now that we have the characterization of a quantum particle in both the gravitational and the physical geometry, 
and have shown that like the conformal transformation, a disformal transformation can be successfully used to 
incorporate the quantum effects in the geometry, one can ask where exactly it differs from the usual picture of 
conformal transformation  and what are the advantages of this identification over the conformal transformation, if any. 
In this section we shall show that a disformal transformation has its advantages, namely it can solve the problems 
addressed in section \ref{ceq}.  
	
\subsection{Continuity equation fixes the disformal transformation completely}\label{dconti}

As we discussed in section \ref{ceq}, the conformal transformation with the quantum mass squared as the 
conformal factor can not give us the right continuity equation. Here we shall show by doing a disformal transformation
of the form eq.(\ref{sdt}), it is possible solve this problem thereby completely fixing both the transformation factors 
$\Omega^{2}$ and $\mathcal{B}$.
	
We start by writing down the disformal version of the eq.(\ref{Cth}) that relates the expansion scalar in both frames 
(see \cite{DK} for the derivation)
\begin{equation}\label{Dth}
\theta^{*}=\mathcal{F}^{-1/2}\Big(\theta+3v^{\alpha}\nabla_{\alpha}\ln\Omega\Big)~.
\end{equation}
Then, as before, we have the analogue of eq.(\ref{cthr}) given by
\begin{equation}
\nabla^{*}_{\mu}\big(\rho v^{*\mu}\big)=\rho\theta^{*}+v^{*\mu}\nabla^{*}_{\mu}\rho=\rho\mathcal{F}^{-1/2}
\Big(\theta+3v^{\alpha}\nabla_{\alpha}\ln\Omega\Big)+\mathcal{F}^{-1/2}v^{\mu}\nabla_{\mu}\rho~.
\end{equation}
Now using eq.(\ref{rhth}) to eliminate $\theta$ from this equation we get 
\begin{equation}
\nabla^{*}_{\mu}\big(\rho v^{*\mu}\big)=\rho\mathcal{F}^{-1/2}\Big(-v^{\alpha}\nabla_{\alpha}\ln M+v^{\alpha}
\nabla_{\alpha}\ln\Omega^{3}\Big)~.
\end{equation}
The requirement that in the disformally transformed frame $\nabla^{*}_{\mu}\big(\rho v^{*\mu}\big)=0$ is  
the correct continuity equation indicates the left hand side to be zero and thus fixes the conformal factor\footnote{$\Omega$ is fixed upto an additive constant which we have taken to be zero} to be 
$\Omega^{3}=M$. 
This, together with the requirement of eq.(\ref{omeB}) also fixes the disformal factor 
$\mathcal{B}=M^{2/3}-M^{2}$.
	
\subsection{Motion of a photon along physical null trajectory}\label{m0}

As was pointed out towards the end of subsection \ref{ceq}, massless particles\footnote{Note that 
by a massless particle, we mean a particle with zero classical mass ($m$), not zero  quantum mass ($M$). 
More precisely what we mean by massless particle here are the particles which moves along the trajectory 
$u_{\mu}u^{\mu}=0$ and hence $p_{\mu}p^{\mu}=0$.} cause a problem in the singularity resolution argument 
because they do not experience any force in frames related by conformal transformations. For the case of 
frames related by disformal transformations (eq.(\ref{sdt}) considered in the previous section), the same problem 
arises because we see from the corresponding relation eq.(\ref{Diacc}), the same conclusion remains valid,
i.e., the extra force term is zero for null trajectories. 
	
We shall argue below that we can cure  this problem in the context of disformal transformation, 
by remembering that, one can make a disformal transformation more general than the one written in eq.(\ref{sdt}), 
where the extra piece need not be necessarily pointing along the direction of the tangent of a particle trajectory. 
This transformation precisely is what given in eq.(\ref{Dist}), where the non conformal piece is along a direction 
specified by the $\phi=const$ hypersurface, which we can choose to be different from  the four velocity. 
This is the advantage of disformal transformations over conformal transformations, namely that one can make 
a massless particle move on a non geodesic motion by going to a frame where square of its four momentum is non 
zero. Below we shall consider a simple form of Maxwell's equation to show clearly how this 
can be done 
(see \cite{CCY} where the transformation of the Maxwell equations  under a disformal transformation are considered by 
assuming the geometrical optics approximation. However we do not make any such approximation).

Let us consider  the motion of a photon in the gravitational geometry (represented by a vector field $A_{\mu}$) 
described by the following Maxwell equations  
\begin{equation}
\nabla^{\mu}\nabla_{\mu}A_{\nu}-\nabla^{\mu}\nabla_{\nu}A_{\mu}=0~, 
\end{equation}
where we  have put all the source terms to zero.  Now, using the Ricci identity which relates the commutator 
of the covariant derivative of a vector field to the Riemann tensor,  
\begin{equation}
\nabla_{\mu}\nabla_{\nu}A^{\alpha}-\nabla_{\nu}\nabla_{\mu}A^{\alpha}=R^{\alpha}_{\beta\mu\nu}A^{\beta}~,
\end{equation}
in the contracted form, and imposing the Lorentz gauge condition $\nabla^{\mu}A_{\mu}=0$, we write the above equation as
\begin{equation}
\nabla^{\mu}\nabla_{\mu}A_{\nu}-R^{\mu}_{\nu}A_{\mu}=0.
\end{equation} 
As before, in a Bohmian treatment, we use the polar from of $A_{\mu}$ 
\begin{equation}
A_{\mu}=\mathcal{C}_{\mu}(x)e^{i\mathcal{S}(x)},
\end{equation}
and substitute in the Maxwell's equation. After separating the real and imaginary parts, and identifying 
$\mathcal{K}_{\mu}=\nabla_{\mu}\mathcal{S}$ as 
the wave vector of the photon, we get the following equations 
\begin{equation}\label{MaB}
g^{\mu\nu}\mathcal{K}_{\mu}\mathcal{K}_{\nu}=\frac{\mathcal{C}^{\nu}}{\mathcal{C}^{2}}
\nabla^{\mu}\nabla_{\mu}\mathcal{C}_{\nu}-R^{\mu}_{\nu}\mathcal{C}_{\mu}\mathcal{C}^{\nu}
\equiv\mathcal{H}(\mathcal{C})~, ~~~ \nabla_{\mu}\big(\mathcal{C}^{2}\nabla^{\mu}\mathcal{S}\big)=0~,
\end{equation}   
where $\mathcal{C}^{2}=g^{\mu\nu}\mathcal{C}_{\mu}\mathcal{C}_{\nu}$ denotes the magnitude of the 
vector $\mathcal{C}_{\mu}$. The first relation gives the magnitude of the wave vector, 
with $\mathcal{H}(\mathcal{C})$ denoting the quantum potential in this case, 
and the second one is essentially the conservation equation.  As can be anticipated from 
eq.(\ref{MaB}), since in this frame, motion of the photon is represented by the non null vector 
$\mathcal{K}_{\mu}$, the motion of photon is along a non geodesic trajectory.
	
Let us now apply the disformal transformation (eq.(\ref{Dist}) with $\alpha=-1$) and see if we can make the 
photon motion a geodesic in the transformed frame. In doing so, the first thing to notice that, since  
the transformation functions in eq.(\ref{Dist}) are real, the phase factor $\mathcal{S}$ and hence the wave vector 
$\mathcal{K}_{\mu}=\nabla_{\mu}\mathcal{S}$ is equal in both the frames. Then, using the inverse  
relation of eq.(\ref{Dist}), the left hand side of first equation of  eq.(\ref{MaB}) reduces to 
\begin{equation}\label{TMaB}
g^{\mu\nu}\mathcal{K}_{\mu}\mathcal{K}_{\nu} \rightarrow \Omega ^{2}g^{*\mu\nu}
\mathcal{K}_{\mu}\mathcal{K}_{\nu}+\frac{\mathcal{B}}
{\Omega^{2}-2X\mathcal{B}}\Big(\phi^{\mu}\mathcal{K}_{\mu}\Big)^{2}~.
\end{equation}
Now it is easy to see by comparing eq.(\ref{MaB}) and eq.(\ref{TMaB}), that if we want to make $\mathcal{K}_{\nu}$ 
null in the transformed frame, then we have to choose the transformation functions such that 
\begin{equation}\label{disconst}
\frac{\mathcal{B}}{\Omega^{2}-2X\mathcal{B}}\big(\phi^{\mu}\mathcal{K}_{\mu}\big)^{2}=\frac{\mathcal{C}^{\nu}}
{\mathcal{C}^{2}}\nabla^{\mu}\nabla_{\mu}\mathcal{C}_{\nu}-R^{\mu}_{\nu}\mathcal{C}_{\mu}\mathcal{C}^{\nu}
=\mathcal{H}(\mathcal{C})~.
\end{equation}
Of course, this single equation does not completely determine the disformal transformation specified by $\Omega,\mathcal{B},\phi^{\mu}$.
We can have another relation from the continuity equation. However unlike the previous case for massive particles, 
these two equations (eq.(\ref{disconst}) and the one obtained from the continuity equation) are not enough for fixing 
both the transformation factors $\Omega$ and $\mathcal{B}$ 
and  the direction of the disformal vector specified by components of $\phi_{\mu}$ uniquely. We  also have to make some choice of 
the factor $\Omega$ (such as pure the disformal transformation with $\Omega= constant$) and/or  of the disformal vector 
(such as timelike ($\phi_{\mu}\phi^{\mu}=-1$) or null disformal transformation ($\phi_{\mu}\phi^{\mu}=0$)). 
The description of the photon's motion  with such explicit choices are left for a future work.
	
The  equation of motion  derived   from eq.(\ref{MaB}) by taking covariant derivative  of both side is the non geodesic equation
\begin{equation}\label{Dng}
\mathcal{K}^{\mu}\nabla_{\mu}\mathcal{K}_{\nu}=\frac{1}{2}\nabla_{\nu}\bigg[\frac{\mathcal{B}}
{\Omega^{2}-2X\mathcal{B}}\big(\phi^{\mu}\mathcal{K}_{\mu}\big)^{2}\bigg]=\frac{1}{2}\nabla_{\nu}\mathcal{H}(\mathcal{C}),
\end{equation}
which, as can be checked, under a disformal transformation satisfying eq.(\ref{disconst}) transforms to the geodesic equation
\begin{equation}\label{Dg}
\mathcal{K}^{*\mu}\nabla^{*}_{\mu}\mathcal{K}^{*}_{\nu}=0.
\end{equation}
Thus the motion of a photon in a gravitational geometry follows a non null trajectory  
(the null trajectories $ds^{2}=g_{\mu\nu}dx^{\mu}dx^{\nu}=0$ in the gravitational geometry are followed by the gravitons \cite{JBe})
and hence its motion is nongeodesic, acted upon by the force due to the quantum potential (see eq.(\ref{MaB}) 
and eq.(\ref{Dng}) respectively). But this motion in a disformally related physical geometry follows a null trajectory,
$ds^{*2}=g^{*}_{\mu\nu}dx^{\mu}dx^{\nu}=0$ of that geometry, where the quantum potential determines 
the required transformation factors. 
As we have shown in section \ref{Ckv}, $p_{\mu}K^{\mu}$ is a conserved quantity along the non 
geodesic motion, provided that $K^{\mu}$ is a conformal Killing vector i.e., a Killing vector of a conformally 
related metric. By using an analogous procedure, it is easy to see that $\mathcal{K}_{\mu}K^{\mu}$ is a conserved 
quantity along the photon trajectory if it satisfies the equation $ \nabla_{(\mu}K_{\nu)}=
-2g_{\mu\nu}K^{\lambda}\nabla_{\lambda}(\ln \mathcal{H}(\mathcal{C}))$, and this quantity should be used 
if one wants to study the deflection of light in gravitational field. 
	
\subsection{Problem of definiteness of mass}\label{pma}

When one uses the standard Klein-Gordon equation to perform a Bohmian treatment,
one gets the resulting quantum Hamilton-Jacobi equation of eq.(\ref{dispersion}). The right side of 
this equation, interpreted as the mass square (denoted as $M^{2}$) is not always positive definite, 
and hence the theory can have tachyonic solutions (see for example \cite{PH}). 
As we have mentioned before, one motivation for transforming to a conformal frame to describe the 
quantum motion of a particle satisfying the Klein-Gordon equation is to avoid this problem, so that in the 
transformed frame the particle has a positive definite mass. But also as pointed out before,  
this comes with other problems such as the wrong continuity equation and the problem 
with massless particles. Most importantly when the quantum mass $M$ is imaginary, 
the conformal factor is itself negative and hence the transformation is not well defined. 

In subsection \ref{dconti}, we have shown that for a massive particle\footnote{For the description of massless 
particle one has to use a more general transformation of eq.(\ref{Dist}). In this subsection, we shall concentrate on 
the case of massive particles only.} by transforming to a disformal frame where two metrics are related by 
eq.(\ref{sdt}), with the transformation factors given in terms of the quantum mass by the relations
\begin{equation}\label{Disfactors}
\Omega^{2}=M^{2/3}, ~~ \mathcal{B}=M^{2/3}-M^{2},~~~ M(x)=\sqrt{\big(m^{2}-f(\mathcal{R})\big)}, 
\end{equation}
one could achieve a consistent  continuity equation for the motion of the quantum particle. 
However the same transformation does not solve the problem of definiteness of mass. 
If the quantum mass is imaginary, then the real root of the conformal factor $\Omega^{2}=M^{2/3}$ in 
eq.(\ref{Disfactors}) becomes negative, so that  the disformal transformation used earlier becomes 
undefined\footnote{In the non-minimally coupled frame, the quantum mass has an extra factor of $\epsilon R$ 
(see eq.(\ref{mass}) above). Can this factor make quantum mass in this frame positive definite? 
Assuming $\epsilon>0$, the answer will  depend on the sign of the Ricci scalar. 
When $R<0$, the $\epsilon R$ term can not make the quantum mass positive definite and  
for  $R>0$ this can make $M$ come out to be positive depending on its relative magnitude 
with the quantum potential. As mentioned in section \ref{minc}, this is one of the advantages of working with 
minimal coupling - the nature of $M$ does not depends on the coupling $\epsilon$.}.
	
In this subsection, we shall show that a possible way out of this problem is  once again to return to the  
general form of the disformal transformation in eq.(\ref{Dist}). However, before doing that let us try to 
locate the problem in the previous setting. When the mass squared is negative, 
we can express the quantum mass formally as $M=|M|e^{i\pi/2}$. Now from the continuity equation
in the transformed frame, we see that the extra phase of $\pi/2$ being  a constant, does not affect this equation 
\begin{equation}\label{inacc}
\nabla^{*}_{\mu}\big(\rho v^{*\mu}\big)=\rho\mathcal{F}^{-1/2}\Big(-v^{\alpha}\nabla_{\alpha}
\ln |M|+v^{\alpha}\nabla_{\alpha}\ln\Omega^{3}\Big)~.
\end{equation} 
Thus by taking $\Omega^{2}=|M|^{2/3}$,  we can still satisfy the continuity equation in the transformed frame,
even when $M$ is imaginary. Looking at the acceleration in eq.(\ref{geo1}) (with $\epsilon=0$), we could similarly 
argue that, by choosing $\sqrt{\mathcal{F}}=\sqrt{\Omega^{2}-\mathcal{B}}=|M|$, this equation can be satisfied 
as well, since the constant phase factor again does not contribute due to the derivative. 
However such a choice does not satisfy the correct dispersion relation $g^{*\mu\nu}p^{*}_{\mu}p^{*}_{\nu}=-1$, 
because under such a transformation, the required relation
\begin{equation}\label{wrodis}
g^{\mu\nu}p_{\mu}p_{\nu}=|M|^{2} ~~ \rightarrow~~~ \bigg[\Omega^{2}g^{*\mu\nu}+\bigg(\frac{\mathcal{B}}
{\Omega^{2}-\mathcal{B}}\bigg)v^{\mu}v^{\nu}\bigg]p_{\mu}p_{\nu}\equiv-\Omega^{2}+\bigg(\frac{-\mathcal{B}|M|^{2}}
{\Omega^{2}-\mathcal{B}}\bigg)\neq|M|^{2}~,
\end{equation}
is not satisfied when we choose $\Omega^{2}-\mathcal{B}=|M|^{2}$. This is the real problem, namely 
to satisfy the correct dispersion relation when quantum mass is imaginary. 
Indeed the Bohmian treatment of Klein-Gordon equation gives the dispersion relation, and not the acceleration
 equation (see eq.(\ref{const})). In the process of taking the covariant derivative, the imaginary factors are ``lost'' in both 
 eq.(\ref{inacc}) and the continuity equation, and one can reach wrong conclusions by considering them. 
 Thus our primary focus  will be on the dispersion relation.
	
We begin the general case by writing above equation of eq.(\ref{wrodis}) for such a transformation in eq.(\ref{Dist}),
\begin{equation}\label{dispe2}
g^{\mu\nu}p_{\mu}p_{\nu}=|M|^{2} ~~ \rightarrow~~~ \bigg[\Omega^{2}g^{*\mu\nu}+\bigg(\frac{\mathcal{B}}
{\Omega^{2}-2\mathcal{B}X}\bigg)\phi^{\mu}\phi^{\nu}\bigg]p_{\mu}p_{\nu}\equiv-\Omega^{2}
+\bigg(\frac{-\mathcal{B}\mathcal{D}^{2}|M|^{2}}
{\Omega^{2}-2\mathcal{B}X}\bigg)=|M|^{2}~,
\end{equation}
where $\mathcal{D}=\phi^{\mu}v_{\mu}$ denotes the projection of the vector $\phi^{\mu}$ along the 
particle 4-velocity.

For our purpose in this subsection, it is sufficient to consider the so called pure disformal transformation, 
where both the conformal and disformal factors are set to a constant (denoted as $\Omega_{0}^{2}$ and $\mathcal{B}_{0}$ 
respectively, with $\Omega_{0}^{2}>0$). Thus  the transformed metric and its inverse are given respectively by
\begin{equation}\label{Dist3}
g^{\star}_{\mu\nu}=\Omega_{0}^{2}g_{\mu\nu}+ \mathcal{B}_{0}\phi_{\mu}\phi_{\nu}~, ~ 
g^{\star\mu\nu}=\Omega_{0}^{-2}\bigg[g^{\mu\nu}- \frac{\mathcal{B}_{0}}
{\Omega_{0}^{2}-2X\mathcal{B}_{0}}\phi^{\mu}\phi^{\nu}\bigg]~, ~ X=-\frac{1}{2}g^{\mu\nu}\phi_{\mu}\phi_{\nu}~.
\end{equation}
All the transformed quantities are denoted by an overhead star. Here the 4-vector $\phi^{\mu}$, 
having  components $(\phi^{0},\phi^{1},\phi^{2},\phi^{3})$, determine the direction of the disformal transformation 
and is to be determined form the correct transformation of the dispersion relation and the continuity equation. 
The above transformation corresponds to a change of proper time $d\tau^{\star2}=\beta^{2}d\tau^{2}$  with 
$\beta=\big(\Omega_{0}^{2}- \mathcal{B}_{0}\mathcal{D}^{2}\big)^{1/2}$.
Then eq.(\ref{dispe2})  gives the first constraint relation to be 
\begin{equation}\label{cost1}
\Omega_{0}^{2}+\bigg(\frac{\mathcal{B}_{0}\mathcal{D}^{2}|M|^{2}}{\Omega_{0}^{2}-2\mathcal{B}_{0}X}\bigg)+|M|^{2}=0~,~\text{with}~,~ 
\mathcal{D}\neq0~~~ \text{unless}~~~|M|=\text{constant}~~.
\end{equation}
The continuity equation gives another constraint. To determine this, we start by writing down the 
transformation rule of the covariant derivative. In the most general case, the change of the Christoffel symbols 
are quite complicated (see for example \cite{CCY, ZGB}), but for our case, it simplifies to the following 
\begin{equation}
\nabla^{\star}_{\mu}v^{\nu}=\nabla_{\mu}v^{\nu}+C^{\mu}_{\nu\alpha}v^{\alpha}~~~\text{where}~~~ 
C^{\mu}_{\nu\alpha}=\bigg(\frac{\mathcal{B}_{0}}{\Omega_{0}^{2}-2X\mathcal{B}_{0}}\bigg)\phi^{\mu}\nabla_{\alpha}\phi_{\nu}~~~.
\end{equation}
Writing the left side in terms of transformed velocity $v^{\star\nu}$, and simplifying, we arrive at
\begin{equation}
\nabla^{\star}_{\mu}v^{\star\nu}=\beta^{-1}\Big(\nabla_{\mu}v^{\nu}+C^{\mu}_{\nu\alpha}v^{\alpha}-v^{\nu}\nabla_{\mu}\beta\Big) ~,
\end{equation}
This gives the desired relation  between the expansion scalars (compare with eq.(\ref{Cth}) and eq.(\ref{Dth}))
\begin{equation}
\theta^{\star}=\beta^{-1}\Big(\theta+C^{\nu}_{\nu\alpha}v^{\alpha}-v^{\nu}\nabla_{\nu}\beta\Big)~.
\end{equation}
The continuity equation can now be expanded to to give
\begin{equation}
\nabla^{\star}_{\mu}\big(\rho v^{\star\mu}\big)=\rho\theta^{\star}+v^{\star\mu}\nabla^{\star}_
{\mu}\rho=\rho\beta^{-1}\Big(\theta+C^{\nu}_
{\nu\alpha}v^{\alpha}-v^{\nu}\nabla_{\nu}\beta\Big)+\beta^{-1}v^{\mu}\nabla_{\mu}\rho~~.
\end{equation}
As before, eliminating $\theta$ from this equation and demanding that continuity equation 
$\nabla^{\star}_{\mu}\big(\rho v^{\star\mu}\big)=0$ is satisfied, we get the second constraint
\begin{equation}\label{cost2}
C^{\nu}_{\nu\alpha}v^{\alpha}-v^{\nu}\nabla_{\nu}\beta-v^{\alpha}\nabla_{\alpha}\ln |M|=0~.
\end{equation}
As before, the pure phase factor does have any effect in the continuity equation, 
so that components of  $\phi^{\mu}$ are real. If we choose the vector field $\phi^{\mu}$ such that, 
given the quantum mass both the constraint eqs. (\ref{cost1}) and (\ref{cost2}) are satisfied, 
then in the transformed frame ($g^{\star}_{\mu\nu}$) the particle has unit mass, thus solving the problem 
of imaginary  quantum mass in the original frame.  Obviously the two constraints of eqs.(\ref{cost1}) and 
(\ref{cost2}) cannot determine all the four components of $\phi_{\mu}$, and we have to make some choices. 

Below we shall briefly sketch such a procedure  when the quantum mass of the particle ($M$) 
moving in the flat spacetime  is imaginary and is a function of the radial coordinate $r$ 
only\footnote{This mass can  corresponds to a stationary state solution of the Klein-Gordon equation 
for which the quantum potential is time independent.}. We  consider the vector field $\phi^{\mu}$  
to be of the form $(1,\phi^{0}(r),0,0)$ and the particle 4 velocity is $v^{\mu}=(1,0,0,0)$ so that $\mathcal{D}=1$. 
Then, as can be checked, the continuity equation is identically satisfied and eq.(\ref{cost1}) gives the 
form of the function $\phi^{0}(r)$ in terms of the quantum mass as 
\begin{equation}\label{phi0}
\phi^{0}(r)=\sqrt{\frac{\Omega_{0}}{\mathcal{B}_{0}}}\sqrt{\frac{\mathcal{B}_{0}-\Omega_{0}-
|M(r)|^{2}}{\Omega_{0}+|M(r)|^{2}}}~~.
\end{equation}
Given the functional form of $|M(r)|$ one can now easily determine the function $\phi^{0}(r)$\footnote{Of course for some 
$|M(r)|$ the function $\phi^{0}(r)$ determined in eq.(\ref{phi0}) may not be real every where. In that case one has 
to take an anastz for the vector field different from the one given here.}.
	
We also mention here that there are ways to make the quantum mass positive definite even in the 
classical background. For example, in \cite{SS}, the authors had shown how to make $M$ positive definite 
by using a different interpretation of Bohmian mechanics and demanding that the mass $M$ 
should have the correct non relativistic limit. In this procedure, they obtained the following formula 
for the quantum  mass ($M$) for  a particle of classical mass $m$
\begin{equation}\label{pqm}
M=m\exp\bigg[\frac{1}{m^{2}}\frac{\Box \mathcal{R}}{\mathcal{R}}\bigg].
\end{equation} 
With this mass formula, one can obtain the correct non relativistic equation of motion of the  
particle as shown in \cite{SS}. The resulting theory has been studied extensively in the context of 
conformal transformations and  curved spacetime (see \cite{RC} for a review)\footnote{This approach was recently 
used in \cite{GRL} to derive a quantum version of the Friedmann equations.}. In this paper, 
we have already shown a way out of this problem by using the disformal transformation, nevertheless it is 
useful to make the theory  free from any tachyonic solution from the start. We leave 
the problem of incorporating the mass of eq.(\ref{pqm}) in our transformation formulas of
eq.(\ref{Disfactors}) for a future study.

\section{Conclusions}\label{sec5}

In the well known approach of incorporating the quantum effects of geometry, conformally related 
spacetimes are used to establish the equivalence between the Bohmian motion of a particle on a
classical background with force free classical motion on a quantum corrected background. 
In terms of this bimetric description of gravity, the Bohmain force can be interpreted as arising due 
to the description of the particle in the gravitational geometry rather than in the physical one, 
and quantum effects are incorporated in the conformal degrees of freedom \cite{SS2}. This line of reasoning can be used to deal with the singularity problem of GR \cite{SD}. 

The first part of this paper is devoted to the discussion of various aspects of this approach. 
In the second part, we have shown how one can  incorporate the quantum nature of a particle 
(both massive and massless) in the  disformal degrees of freedom of the physical geometry. 
The conformal transformation is equivalent to a uniform coordinate dependent scaling in every 
spacetime direction. On the other hand, in a disformal transformation of the form given in eq.(\ref{Dist}), 
we not only do a uniform scaling in every direction (conformal part), we also scale a particular direction 
(chosen by the normal vector $\phi_{\mu}$) differently from other directions. This fact is more transparent 
from the so called  pure disformal transformation, where we only scale the direction chosen by the velocity 
vector $v_{\mu}$, and the other directions (perpendicular to $v_{\mu}$ in an orthonormal coordinate frame) 
do not scale at all i.e., the conformal factor is just a constant. 

What we have shown in this paper imply that the quantum effects can can viewed as a disformal transformation  
between the gravitational and the physical geometry, i.e., if we consider the motion of quantum particle 
in gravitational geometry in presence of the quantum potential, this is equivalent to the classical free fall motion 
in  a  disformally transformed spacetime, where the transformation is done  along the direction chosen 
by the  the particle four velocity. The quantum force arises in the gravitational  geometry only because we 
had chosen the ``wrong" frame for the analysis.  Most importantly, by using disformal transformation  
we have shown here that one can solve the usual problems in the conformal version, such as wrong continuity 
equation and the problem of definiteness of mass.

Before concluding we mention here another future application of the formalism constructed here. We notice that the 
second factor in the transformation relation of expansion scalar derived in eq.(\ref{Cth}) indicates that it is not a 
conformally invariant quantity. Similarly it is easy to show, taking symmetric and antisymmetric part of eq.(\ref{4tr}) 
respectively that, the  shear and rotation tensors are also not conformally invariant. Applying directional derivative to  
eq.(\ref{Cth}) we see that Raychaudhuri equation is also not conformally invariant 
\begin{equation}
\frac{d\tilde{\theta}}{d\tilde{\tau}}=\Omega^{-2}\bigg[\frac{d\theta}{d\tau}-\Omega^{-1}\bigg(\theta+3\frac{d\ln \Omega}
{d\tau}\bigg)\frac{d\Omega}{d\tau}+3\frac{d^{2}\ln\Omega}{d\tau^{2}}\bigg].
\end{equation}
In GR, such quantities are problematic because  for a particle trajectory these are the observable quantities 
and appear in the  Raychudhuri equation in their scalar form, so these should be conformally invariant. 
The root of this problem lies  in the  transformation relation of Christoffel symbols eq.(\ref{Chtr}) and the 
fact that in GR Christoffel connections are symmetric in the lower indices. This makes geodesic equation 
to changes differently in conformally related frames. This problem has recently been addressed in \cite{LP}, 
where the authors have shown how to make the geodesic equation invariant  under a conformal transformation 
in the framework of Einstein-Cartan gravity, where the presence of non zero torsion makes this possible.  
It will be  interesting to see how such modifications can affect  the quantum motion of a particle in curved 
spacetime along the lines discussed in this paper.

\end{document}